\documentstyle[prl,aps]{revtex}
\input epsf
\begin{document}
\draft
\twocolumn

{\noindent \bf Comment on ``Nonclassical Smoothening of Nanoscale
Surface Corrugations''} \vspace*{0.3cm}

In a recent Letter \cite{Nonclassical_prl} Erlebacher et al.\ describe
the experimental observation of nonclassical smoothening of a
crystalline surface. In their experiment, ripples of typical
wavelengths $\lambda_x=290-550nm$ in the $x$-direction were formed on
Si(001) by sputter rippling and then annealed at $650-750^\circ$C. They
report that in contrast with the classical exponential decay with time,
the ripple amplitude $A(t)$ followed an inverse linear decay of the
type $A(t)=A(0)/[1+\alpha t]$. Erlebacher et al.\ interpret their
results in terms of a one dimensional step flow model, and compare them
with theoretical work of Ozdemir and Zangwill (OZ)
\cite{OzdemirZangwill}. While we have no reservations about the
experimental results, we claim that they are {\em inconsistent} with
the one dimensional step flow model, and a fully two-dimensional model
is necessary in order to describe the behavior of the experimental
system.
\begin{figure}[h]
\centerline{ \epsfxsize=80mm \epsffile{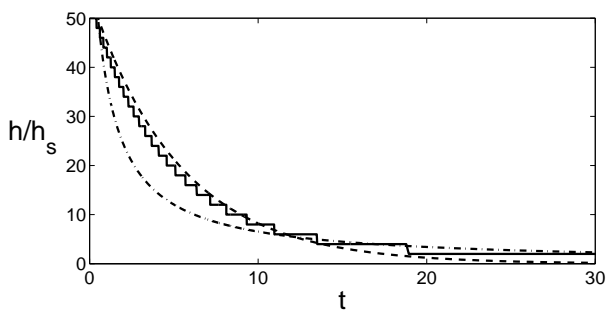}}
\vspace*{2mm}\caption{Amplitude relaxation of an initially sinusoidal
grating (solid line) measured in units of the height of an atomic step,
$h_s$. The dashed and dash-dotted lines are fits to an exponential
decay and an inverse linear decay, respectively. Time is in arbitrary
units.} \label{figure1}
\end{figure}

The mass transport rate limiting process for surface evolution depends
on the typical terrace width, $\Delta x$, and on the length
$\ell=D_s/k$, where $D_s$ is the adatom surface diffusion constant and
$k$ is the rate of attachment of adatoms to step edges. The kinetics of
the system is diffusion limited (DL) when $\ell\ll\Delta x$ and
attachment-detachment limited (ADL) when $\ell\gg\Delta x$. Bartelt et
al.\ showed \cite{Bartelt} that for Si(001) surfaces, $\ell\geq 500nm$
for temperatures in the range $670-1010^\circ$C. This value of $\ell$
is more than an order of magnitude larger than typical terrace widths
in the experiment of Erlebacher et al., and therefore their system
should exhibit ADL kinetics.

We studied the relaxation of large-amplitude one dimensional gratings
\cite{sine_paper} in terms of the same one dimensional step flow model
used by OZ \cite{OzdemirZangwill}. We found that unless the system is
purely DL, the height of the grating decays {\em exponentially} in
time, agreeing entirely with the experiments of Keeffe et al.\
\cite{Keeffe}. Figure \ref{figure1} shows that even when
$\ell=\lambda_x$ and the grating amplitude is 50 steps (values comparable to the
experimental ones), an exponential decay fits the step
model results better than an inverse linear decay. Thus, the
experimental results of Erlebacher et al.\ are {\em inconsistent} with
the one dimensional theory.

The confusion arises because of the inaccurate statement made by OZ,
claiming to have found inverse linear decay in all kinetic limits, but
presenting numerical evidence only for DL kinetics. Our work
\cite{sine_paper} shows unequivocally that the decay is exponential in
all cases but purely DL kinetics (which shows inverse linear
relaxation).

We now argue that a two dimensional model is needed in order to
describe the experiment of Erlebacher et al. The atomic force
microscope images in \cite{Nonclassical_prl} indicate that the
corrugations are not purely one dimensional. In addition to the short
wavelength ripples along the $x$-direction, the surface is corrugated
in the transverse $y$-direction with a typical wavelength $\lambda_y
\approx 10\, \lambda_x$. We showed \cite{cone_paper} that for radially
symmetric morphology the diffusion current $J$ satisfies $J \propto
d\mu/d\zeta$, where $\mu$ is the local step chemical potential and
$\zeta=r+2\ell\, h(r)/h_s$, with $h_s$ being the height of a step. The
generalization of this relation for non radially symmetric morphology
allows an estimate of the order of magnitude of the peak-to-valley
directional currents $J_x$ and $J_y$ according to $J_i\propto \Delta
\mu/(\lambda_i/2+2\ell\Delta h/h_s)$ for $i=x$ or $y$. Here $\Delta
\mu$ is the peak-to-valley chemical potential difference and $\Delta h$
is the peak-to-valley height difference. According to this, $J_y/J_x
\approx 0.95$ in the experiments of Erlebacher et al. A one dimensional
model is appropriate only when $J_y$ is negligible compared with $J_x$,
which is clearly not the case here.

\vspace*{0.5cm} {\noindent Navot Israeli$^1$ and Daniel
Kandel$^2$}\\
\hspace*{5pt} $^1$Department of Physics\\
\hspace*{10pt} University of Illinois at Urbana-Champaign\\
\hspace*{10pt} 1110 West Green Street\\
\hspace*{10pt} Urbana, Illinois 61801-3080\\
\hspace*{5pt} $^2$Department of Physics of Complex Systems\\
\hspace*{10pt} Weizmann Institute of Science\\
\hspace*{10pt} Rehovot 76100, Israel

\vspace*{0.2cm}{\noindent 


\begin{thebibliography}{99}
\vspace{-1.5cm}
\bibitem{Nonclassical_prl}
J. Erlebacher, M. J. Aziz, E. Chason, M. B. Sinclair and J. A. Floro,
Phys. Rev. Lett. {\bf 84}, 5800 (2000).

\bibitem{OzdemirZangwill}
M. Ozdemir and A. Zangwill, Phys. Rev. B {\bf 42}, 5013 (1990).

\bibitem{Bartelt}
N. C. Bartelt, W. Theis and R. M. Tromp, Phys. Rev.\ B {\bf 54}, 11741
(1996).

\bibitem{sine_paper}
N. Israeli and D. Kandel, Phys. Rev. B {\bf 62}, 13707 (2000).

\bibitem{Keeffe}
M. E Keeffe, C. C. Umbach and J. M. Blakely, J. Phys. Chem. Solids {\bf
55}, 965 (1994).

\bibitem{cone_paper}
N. Israeli and D. Kandel, Phys. Rev. B {\bf 60}, 5946 (1999).
\end{thebibliography}
\end{document}